\documentclass[aps,pre,twocolumn,showpacs,amsmath,amssymb,amsfonts]{revtex4}
\usepackage{epsfig}
\usepackage{graphicx}
\usepackage{dcolumn}\usepackage{braket}
\usepackage{bm}
\usepackage{soul, xcolor}
\usepackage{xcolor,cancel}
\usepackage[titletoc,title]{appendix}
\usepackage{subfig}
\usepackage{floatrow}

\usepackage{color}
\usepackage[normalem]{ulem} %To strikeout text

\begin{document}
\setstcolor{red}
\title{Role of the sonic scale in the growth of magnetic field in compressible turbulence}

\author{Itzhak Fouxon}\email{itzhak8@gmail.com}
\author{Michael Mond}\email{mondmichael@gmail.com}
%\affiliation{$^1$ {\color{red}{you can use now BGU affiliation}}}
\affiliation{Department of Mechanical Engineering, Ben-Gurion University of the Negev,
P.O. Box 653, Beer-Sheva 84105, Israel}

\begin{abstract}

We study the growth of small fluctuations of magnetic field in supersonic turbulence, the small-scale dynamo. The growth is due to the fastest turbulent eddies above the resistive scale. We observe that for supersonic turbulence these eddies are effectively incompressible which creates a robust structure of the growth. The eddies are localised below
the sonic scale $l_s$ defined as the scale where the typical velocity of the turbulent eddies equals the speed of sound. Thus the flow below $l_s$ is effectively incompressible and the field growth proceeds as in incompressible flow. At large Mach numbers $l_s$ is much smaller than the integral scale of turbulence so the fastest growing mode of the magnetic field belongs to small-scale turbulence.
We derive this mode and the associated growth rate numerically in a white noise in time model of turbulence. The relevance of this model relies on considering evolution time larger than the correlation time of turbulence.

\end{abstract}
\maketitle

{\it Introduction}. The problem of the growth of small fluctuations of magnetic field in turbulent flows of a conducting fluid is one of the most fundamental problems of magnetohydrodynamics (MHD) \cite{ll8,xl}. Here the back reaction of the magnetic field on the flow is neglected due to the field's smallness. Growth occurs when the exact solution of the MHD equations with zero magnetic field is unstable. The field's growth brings a non-trivial state where the inertial and magnetic degrees of freedom interact strongly. Oppositely, if the field decays then the steady state is a purely inertial turbulence.

The growth has most significant consequences in astrophysical context \cite{xl}. There turbulence is often
characterised by a large Mach number $Ma$ (ratio of the rms velocity and the speed of sound) implying a strongly compressible flow. Previous studies of magnetic field turbulent growth have been carried out mainly for
incompressible turbulence with small Mach numbers, see e. g. \cite{xl,kas,rs,ld,krs,verg,cfkv,dv,horvarp,isz}. In this Letter we consider the growth in the compressible homogeneous isotropic turbulence. This is a paradigm problem for the study of the growth in non-helical random flow, the so-called fluctuation or small-scale dynamo \cite{xl}. We observe that the growth is a small-scale phenomenon. The characteristic time of the growth at a given scale is the eddy turnover time at this scale. We assume that this time decreases with the scale which seems inevitable property of turbulence for any Mach number.
Then the overall growth is dominated by the smallest possible eddies that must yet be larger than the resistivity scale $l_{res}$ so that the growth is not inhibited by the resistivity.
Thus the fastest growing mode is localised near $l_{res}$. Model calculations indicate that the mode's decay beyond $l_{res}$ is fast obeying a stretched exponential law \cite{dv}. We demonstrate here that this property must hold also in the Navier-Stokes (NS) turbulence. The corresponding growth rate is the inverse characteristic time-scale of turbulence at $l_{res}$, a fact well-known in the incompressible case \cite{kas,rs,dv,xl} that carries over to compressible turbulence. The local Mach number is usually small at $l_{res}$ so that the flow compressibility, possibly large at large scales, is irrelevant and the growth proceeds as in the incompressible case. This observation is reinforced by the fact that compressibility, that is relevant above the sonic scale \cite{sonic}, decreases the growth rate of the magnetic field \cite{rk,schk,bov,vvk,fos} thus further decreasing the relevance of large scales.

The consideration above relies on the assumption that the Mach number at $l_{res}$ is small which holds in practice typically. In addition, $l_{res}$ must be larger than the smallest (viscous) scale of turbulent fluctuations which is the case of the small magnetic Prandtl number \cite{xl}. A similar consideration for large Prandtl number case \cite{cfkv,isz,schk} will be provided elsewhere%\cite{current}
. Previous studies of the magnetic dynamo in compressible turbulence \cite{krs,rk,bov,vvk} did not consider the sonic scale and the incompressibility of the turbulence at $l_{res}$ as implied by the typically large value of $l_s/l_{res}$.

{\it The Kraichnan-Kazantsev Model.} Turbulence leads to the magnetic field's growth by stretching the magnetic field lines. However it also produces the opposite effect of mixing the lines and creating sharp field contrasts. Below the small resistive scale $l_{res}$ these contrasts are damped efficiently by the resistivity. Thus turbulence could both cause the increase and the decrease of the magnetic field. In fact, turbulence accelerates the field's damping in the case of planar flows \cite{ll8}. In contrast, in the three-dimensional case turbulent stretching dominates so that the incompressible homogeneous isotropic turbulence produces exponential growth of small fluctuations of the magnetic field. This result was first obtained theoretically by Kazantsev using a model of turbulence \cite{kas}. The flow is considered as a Gaussian random field with zero mean and white in time pair correlations that obey scaling in space. It was demonstrated that the field grows if the scaling exponent fits the Kolmogorov scaling.  Since a similar model was introduced independently by Kraichnan for the study of passive scalar transport by incompressible turbulence \cite{kr} it is called Kazantsev-Kraichnan (KK) model \cite{dv,fos}. The KK model is not a good model of the Navier-Stokes (NS) turbulence which is not a white noise in time. However we demonstrate that on large time scales, that are of interest in the dynamo problem, turbulence does look as a white noise due to a Langevin-type description of the long-time evolution. The KK model then provides the simplest approximation for the effective evolution operator.

One can hope that only crude properties of turbulence determine its effect on the magnetic field and that these properties are captured by the KK model. In fact numerical simulations of the NS turbulence \cite{frs} confirm the qualitative validity of the KK model predictions in the incompressible case, see \cite{verg,horvarp} and review \cite{xl} for discussion and further references. The model's conclusion that the dynamo is a small-scale phenomenon, which we derive here in the compressible case, seems a robust fact. Once this is accepted the growth in the NS turbulence is driven by the effectively incompressible turbulent eddies below $l_s$. The potential component of the flow is also present at these scales - there are shock waves whose dissipative scale is typically much smaller than $l_s$. However this component is small as confirmed for instance by the lack of density fluctuations below $l_s$, see e. g. \cite{dust}.

The evolution of the magnetic field $\bm B$ obeys \cite{ll8},
\begin{eqnarray}&&\!\!\!\!\!\!\!\!\!\!\!\!\!\!
\partial_t\bm B+(\bm v\cdot \nabla) \bm B=(\bm B\cdot \nabla) \bm v-\bm B\nabla\cdot \bm v+\eta\nabla^2 \bm B, \label{magn}
\end{eqnarray}
where $\bm v$ is the turbulent flow and $\eta$ is the magnetic resistivity. This equation is linear in $\bm B$ and the assumption of the field smallness tells that the field's back reaction on the flow, given by the Lorentz force in the NS equations, may be neglected. Thus $\bm v$ is considered as a given flow independent of $\bm B$. We observe that if we introduce a linear time-dependent operator ${\hat L}(t)$ by,
\begin{eqnarray}&&\!\!\!\!\!\!\!\!\!\!\!\!\!\!
{\hat L}(t)\bm B=-(\bm v\cdot \nabla) \bm B+(\bm B\cdot \nabla) \bm v-\bm B\nabla\cdot \bm v+\eta\nabla^2 \bm B,
\end{eqnarray}
then we can write a formal solution for $\bm B$ as,
\begin{eqnarray}&&\!\!\!\!\!\!\!\!\!\!\!\!\!\!
B_i(t, \bm x_1)=\int {\hat W}_{il}(t, \bm x_1, \bm x_1')B_{0l}(\bm x_1') d\bm x_1', \label{fd11}
\end{eqnarray}
where $\bm B_0(\bm x)$ is the initial condition and we introduced the time-ordered exponent of ${\hat L}(t)$,
\begin{eqnarray}&&\!\!\!\!\!\!\!\!\!\!\!\!\!\!
{\hat W}(t)\equiv T \exp\left(\int_0^t {\hat L}(t')dt'\right).
\end{eqnarray}
We are interested in the long-time behavior of the correlation tensor $T_{ik}(t, \bm r)\equiv \left\langle B_i(t, \bm x_1)B_k(t, \bm x_2)\right\rangle$. Here $\bm r=\bm x_2-\bm x_1$ and spatial averaging is designated
by angular brackets. Multiplying Eq.~(\ref{fd11}) by $B_k(t)$ and averaging
%,
%\begin{eqnarray}&&\!\!\!\!\!\!\!\!\!\!\!\!\!\!
%\int  {\hat W}_{il}(t, \bm x_1, \bm x_1'){\hat W}_{kr}(t, \bm x_2, \bm x_2') B_{0l}(\bm x_1')B_{0r}(\bm x_1') d\bm x_1'd\bm x_2'
%\end{eqnarray}
we obtain,
\begin{eqnarray}&&\!\!\!\!\!\!\!\!\!\!\!\!\!\!2
T_{ik}(t, \bm r)=\int M_{ik, lr}(t, \bm r, \bm r')T_{lr}(\bm r')d\bm r'. \label{integral}
\end{eqnarray}
We made the usual assumption that the statistics of $\bm B_0(\bm x)$ is independent of the flow (insignificant for the long-time asymptotic behavior) and defined the matrix element by averaging over $\bm x_1$,
\begin{eqnarray}&&\!\!\!\!\!\!\!\!\!\!\!\!\!\!
M_{ik, lr}(t, \bm r, \bm r')\equiv \left\langle {\hat W}_{il}(t, \bm x_1, \bm x_1'){\hat W}_{kr}(t, \bm x_1+\bm r, \bm x_2')\right\rangle, \label{mal}
\end{eqnarray}
where $\bm r'=\bm x'_2-\bm x'_1$ and spatial homogeneity is used. We observe that at times much larger than the turbulence correlation time, given by the eddy turnover time $t_L$ of the large-scale eddies with scale $L$, we can introduce a time-independent operator ${\hat L}_{eff}$ so that,
\begin{eqnarray}&&\!\!\!\!\!\!\!\!\!\!\!\!\!\!
M_{ik, lr}(t, \bm r, \bm r')=\exp\left(t{\hat L}_{eff}\right)_{ik, lr}(\bm r, \bm r'), \label{esp}
\end{eqnarray}
where ergodicity has been assumed, according to which spatial average is equal to the average over velocity ensemble. The reason for this exponential property is that at these times ${\hat W}(t)$ is the product of a large number $\sim t/t_L$ of independent operators of evolution over time intervals of order $t_L$. Thus the average in Eq.~(\ref{mal}) has asymptotic behavior of a product of $\sim t/t_L$ independent terms implying exponential dependence on $t$, cf. with the finite-dimensional case of the Jacobi matrix e. g. in \cite{review}.

We conclude that as long as we are interested in the long-time properties of the evolution of a magnetic field in a turbulent flow the pair-correlation obeys,
\begin{eqnarray}&&\!\!\!\!\!\!\!\!\!\!\!\!\!\!
\partial_t T_{ik}=({\hat L}_{eff})_{ik, lr}T_{lr}=\int {\tilde K}_{ik, lr}(\bm r, \bm r')T_{lr}(\bm r')d\bm r', \label{rk}
\end{eqnarray}
where we introduced the kernel ${\tilde K}$ of ${\hat L}$, cf. Eqs.~(\ref{integral}) and (\ref{esp}). The operator ${\hat L}_{eff}$ integrates global spatio-temporal action of turbulence on the magnetic field and can hardly be derived from the flow. This operator is non-local in space and its action on a function that decays fast beyond a scale $r$ results in a function that decays fast at a scale larger however comparable with $r$. This is because the operator ${\hat L}_{eff}$ forms at scale $r$ during the correlation time of eddies at this scale, when the distances between fluid particles change by a factor of order one \cite{frisch}. Thus modelling of ${\tilde K}_{ik, lr}(\bm r, \bm r')$ is necessary, cf. \cite{fm}. The simplest way to introduce a consistent model with first order in time evolution of $T_{ik}$ is by using in Eq.~(\ref{magn}) instead of the turbulent flow $\bm v(t, \bm x)$ an artificial flow $\bm u(t, \bm x)$ which is a white noise in time. The flow is Gaussian and it has zero mean and the pair-correlation function \cite{arxiv,phase,fm,fos},
\begin{eqnarray}&&\!\!\!\!\!\!\!\!\!\!\!\!\!\!\!\!\!
\langle v_i(t_1, \bm x_1)v_k(t_2, \bm x_2)\rangle=\delta(t_2-t_1)\left[V_0\delta_{ik}-K_{ik}(\bm r)\right],\label{mdl}
\end{eqnarray}
where $K_{ik}$ is defined by (observe that $K_{ik}(r=0)=0$),
\begin{eqnarray}&&\!\!\!\!\!\!\!\!\!\!\!
2K_{ik}=\left[\frac{(r^4u)'}{r^3}-c\right]r^2\delta_{ik}-\left[\frac{(r^2u)'}{r}-c\right]r_ir_k, \label{kdf}
\end{eqnarray}
where $u(r)$ and $c(r)$ are certain functions of $r$. The model assumes spatial homogeneity and isotropy. For incompressible flow $c\equiv 0$. It can be demonstrated that in this model $T_{ik}(t, \bm r)$ obeys closed equation,
\begin{eqnarray}&&\!\!\!\!\!\!\!\!\!\!\!\!\!\!
\partial_t T_{ik}\!=\!\nabla_{n}\nabla_l \left(K_{nl} T_{ik}\!+\!K_{ik} T_{nl}\!-\!K_{nk} T_{il}\!-\!K_{il} T_{nk}\right)
\nonumber\\&&\!\!\!\!\!\!\!\!\!\!\!\!\!\!
+2\eta \nabla^2 T_{ik}\equiv({\hat L}_{eff}^{KK})_{ik, lr}T_{lr}. \label{tensor0}
\end{eqnarray}
Since the equation in this form is seemingly missing in the literature (see \cite{horvarp} for incompressible case) then we provide the rather lengthy derivation in the SI, cf. \cite{xl}.
The equation gives the KK model of ${\hat L}_{eff}$ that describes ${\tilde K}(\bm r, \bm r')$ in Eq.~(\ref{rk}) as a local operator proportional to $\delta(\bm r-\bm r')$ and its first two derivatives.
Considering ${\hat L}_{eff}$ as an infinite series of derivatives of different orders, the KK model provides a consistent way of cutting off the series at the operators of second order
(by Pawula theorem cutting this series at any higher finite order can be problematic \cite{risken,fm}). The KK model is the spatially local model of ${\hat L}_{eff}$ that satisfies the statistical symmetries. The model must describe well the large-scale behavior of the fastest growing mode localized at scale $l_{res}$ described below. Thus the conclusion below on the stretched exponential decay
of the fastest growing mode at scales much larger than $l_{res}$, obtained within the KK model is highly plausible to apply to the NS turbulence.

The equation for the tensor $T_{ik}$ can be reduced to a scalar equation by introducing the longitudinal, $M_L(r)$, and transversal, $M_N(r)$, correlation functions \cite{kas,xl,fos},
\begin{eqnarray}&&\!\!\!\!\!\!\!\!\!\!\!\!\!\!
T_{ik}\!=\!\left(\delta_{ik}-{\hat r}_i{\hat r}_k\right) M_N+{\hat r}_i{\hat r}_k M_L,\ \ M_L={\hat r}_i{\hat r}_k T_{ik}, \label{sl}
\end{eqnarray}
where ${\hat r}_i=r_i/r$. We have by solenoidality of $\bm B$ that $\nabla_kT_{ik}=0$ that as can be readily checked gives \cite{xl},
\begin{eqnarray}&&\!\!\!\!\!\!\!\!\!\!\!\!\!\!
M_N\!=\!(r^2M_L)'/(2r)=M_L+rM_L'/2. \label{s}
\end{eqnarray}
Thus $T_{ik}$ grows if $M_L$ does. Introducing,
\begin{eqnarray}&&\!\!\!\!\!\!\!\!\!\!\!\!\!\!\!\!\!
K_{ik}(\bm r)=S_N(r)\delta_{ik}+{\hat r}_i{\hat r}_k (S_L(r)-S_N(r)), \label{dfkl}
\end{eqnarray}
we find by calculation provided in SI that $M_L(t, r)$ obeys,
\begin{eqnarray}&&\!\!\!\!\!\!\!\!\!\!\!\!\!\!
\partial_t M_L\!=\!(2\eta+S_L)\partial_r^2 M_L+\left(S_L'+\frac{4S_L}{r}
+\frac{8\eta}{r} \right)\partial_r M_L\nonumber\\&&\!\!\!\!\!\!\!\!
+\left(S_N'+S_L'+\frac{S_L-S_N}{r}\right)\frac{2M_L}{r}.\label{fd111}
\end{eqnarray}
We have by comparison of Eqs.~(\ref{kdf}) and (\ref{dfkl}),
\begin{eqnarray}&&\!\!\!\!\!\!\!\!\!\!\!
2S_N=4r^2u+r^3u'-cr^2,\ \ S_L=r^2u. \label{sls}
\end{eqnarray}
This gives in terms of $c(r)$ and $u(r)$,
\begin{eqnarray}&&\!\!\!\!\!\!\!\!
\partial_t M_L\!=\!(2\eta+r^2u)\partial_r^2 M_L+\left((r^2u)'+4ru
+\frac{8\eta}{r} \right)\partial_r M_L
\nonumber\\&&\!\!\!\!\!\!\!\!
+\left(r^2u''+8ru'+10u-(r c)'\right)M_L,\label{fd}
\end{eqnarray}
so that the compressible component $c$ changes only the last term in the equation. Since $(r c)'>0$ then at fixed $u$ roughly this term decreases the growth of the magnetic field, cf. above.

{\it The Schr{\"o}dinger Equation.} Following Kazantsev \cite{kas} and \cite{fos}, the substitution
\begin{equation}
M_L(r,t)={\hat \psi (r,t)} r^{-2}[2\eta+S_L(r)]^{-1/2},
\end{equation}
transforms Eq.~(\ref{fd}) into the following imaginary-time Schr{\"o}dinger equation with space-dependent mass for the new dependent variable
${\hat \psi (r,t)}$:
\begin{equation}
-\partial _t{\hat \psi}=-\frac{1}{m(r)}\partial^2_r {\hat \psi}+U(r){\hat \psi},
\label{schrodinger}
\end{equation}
where $m(r)=1/[2\eta +S_L(r)]$ and
\begin{equation}
U(r)=\frac{\partial ^2_r S_L}{2}-\frac{[\partial _r S_L]^2}{4[2\eta+S_L]}+\frac{2[2\eta+S_N-r\partial _r S_N]}{r^2}.
\end{equation}
Further decomposition of the solution into ${\hat \psi (r,t)}=\psi (r)e^{-\gamma t}$ results in an eigenvalue problem of the Shr{\"o}dinger operator on the right hand side of Eq.~(\ref{schrodinger}). The fastest growing (in the no dynamo case not considered here \cite{kas} the slowest decaying) mode is given by the ground state of this operator.
%For the study as can be seen from the variational principle \cite{fos}.
%us, negative eigenvalues $\gamma <0$ are associated with the exponential growth in time of the magnetic correlation whose spatial structure is given by the eigenstate $\psi (r)$..

We assume that the ratio of the integral scale of turbulence $L$ to $l_{res}$ is large so that scales beyond $L$ are irrelevant and we can continue the scaling at $r<L$ to $L=\infty$ as in \cite{kas}, see \cite{dv,horvarp}. Similarly the viscous scale, below which the flow is smooth, is assumed to be smaller than $l_{res}$ allowing to disregard that scaling range \cite{kas,dv,horvarp}. The behavior of magnetic
field at ${\rm Ma}\ll 1$ is as in the incompressible case so we consider ${\rm Ma}\gtrsim 1$.

{\it Model for large ${\rm Ma}$.} The difference of the scaling exponents of the solenoidal and potential components of the flow at $l_s\lesssim r\lesssim L_s$ decreases with ${\rm Ma}$.
In the limit of large ${\rm Ma}$, starting at about ${\rm Ma}\simeq 6$ the exponents are close and $l_s\ll L$, see \cite{krit2007} (dependence on compressibility of the stirring force \cite{fedor} can be disregarded for the robust phenomenon considered here. At ${\rm Ma}\ll 1$ the flow's components scale differently, see e. g. \cite{gotoh}). Thus ${\rm Ma}\gtrsim 6$ can be modeled as, see Eq.~(\ref{sls}) and \cite{phase,fos,fm},
\begin{eqnarray}&&\!\!\!\!\!\!\!\!
S_L(r)\!=\!2Dr^{\xi}(1\!+\!{\cal C}\xi),\ \ S_N(r)\!=\!Dr^{\xi}(2\!+\!\xi(1\!-\!{\cal C})), \label{us}
\end{eqnarray}
for $l_s\lesssim r\lesssim L_s$. Here $0\le{\cal C}\le1$ denotes the compressibility ratio such that for ${\cal C}=0$ the flow is solenoidal while for ${\cal C}=1$ it is potential. The constant $D$ characterises the fluctuation's strength. The scaling exponent $\xi$ is given by $\xi=(1+a)/2$ where $a$ is the decay exponent of the spherically normalized spectrum $k^{-a}$, see \cite{fm,review}. Thus $\xi=4/3$ describes the Kolmogorov scaling.
In the limit of large ${\rm Ma}$ the common Burgers scaling of the components $k^{-2}$ is a reasonable hypothesis \cite{bls} that does not contradict observations \cite{krit2007} and gives $\xi=3/2$. Here and below slight changes in the exponents' values do not change the conclusions.

{\it Effect of the Sonic Scale.} As discussed above, the growth of the magnetic field is driven effectively by incompressible eddies below the sonic scale $l_s$, where the flow is basically solenoidal namely characterised by compressibility ratio ${\cal C}=0$. Therefore, it is plausible to assume that ${\cal C}$ depends on the eddies length scale. In order to investigate the effect of such dependence on the growth of the magnetic correlation the following scale-dependent model is examined:
\begin{eqnarray}&&
{\cal \hat {C}} (r)={\cal C}\tanh \Bigl (\frac{r}{l_s}\Bigr ),\ \ {\cal \hat {\xi}}(r)=\xi_0+(\xi-\xi_0)\tanh \Bigl (\frac{r}{l_s}\Bigr ),\nonumber\\&&
{\cal \hat {D}}(r)r^{\hat {\xi}(r)}=D_0r^{\xi _0}+(Dr^{\xi}-D_0r^{\xi _0})\tanh \Bigl (\frac{r}{l_s}\Bigr ),
\label{tanh}
\end{eqnarray}
such that below the sonic scale the compressibility ratio is practically zero while for scales well above the sonic scale it approaches the compressibility ratio that is associated with the entire system.
Similarly the scaling exponent of the solenoidal component changes from the incompressible value $\xi_0$ at $r\ll l_s$ to $\xi$ at $r\gg l_s$. We can use the Kolmogorov value of $\xi_0=4/3$ which is well within the range $\xi>1$ where the field grows \cite{kas,verg} so slight intermittency deviations from $4/3$ are of no qualitative significance. Finally we require that the scaling laws at $r<l_s$ and $r>l_s$ agree at $r=l_s$ by order of magnitude setting $D_0 l_s^{\xi_0}=Dl_s^{\xi}$. %We observe that the scaling in the supersonic range is smoother.

\begin{figure}
  \centerline{\includegraphics[width=7cm, height=5cm]{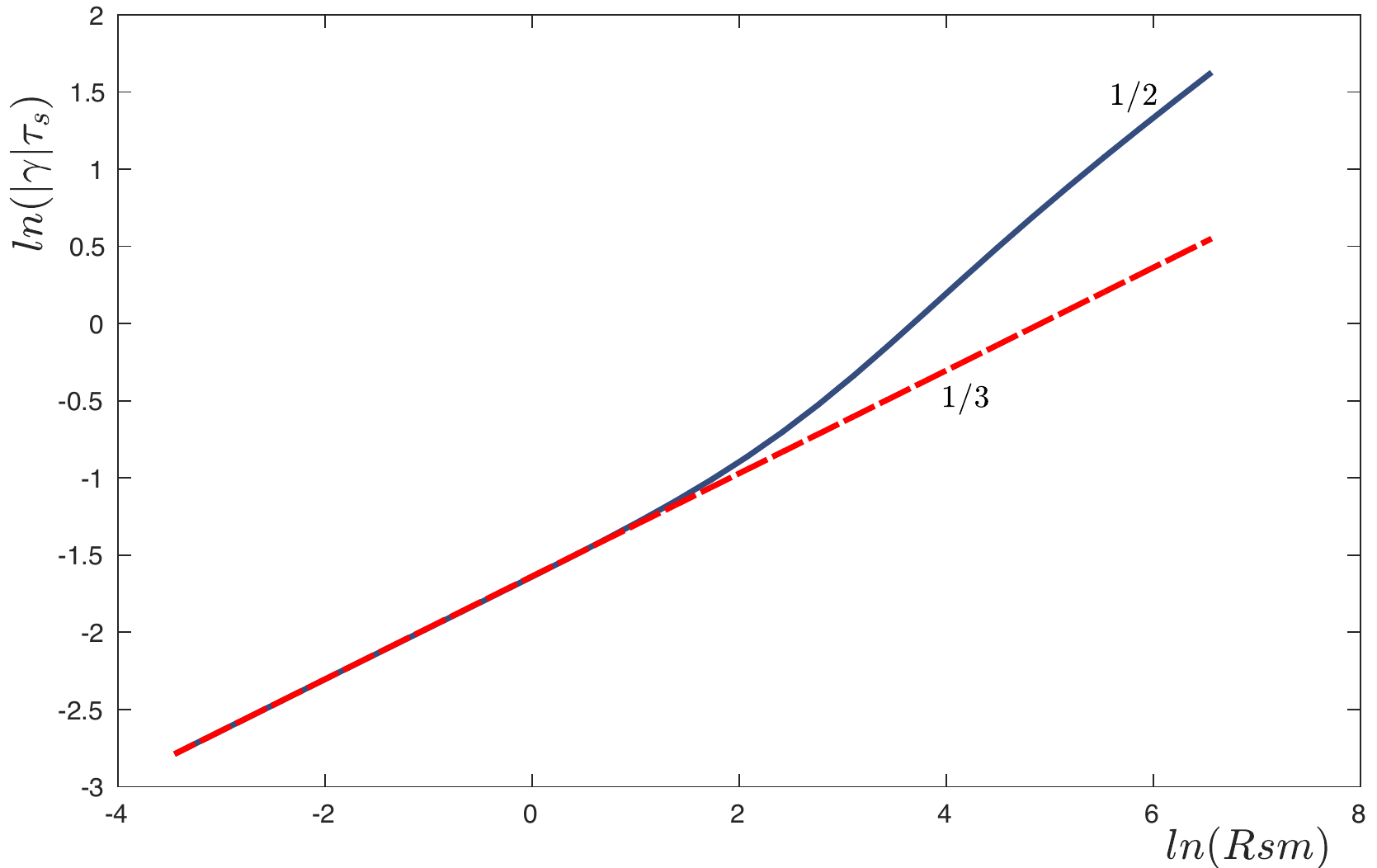}}% Images in 100% size
  \caption{The magnetic field growth rate as a function of the sonic magnetic Reynolds number for ${\cal C}=0.8$. If $\xi_0$ were equal to $\xi$ then there would still be an increase of the growth rate which would be displaced upwards keeping the same slope. }
\label{fig:gamma}
\end{figure}

\begin{figure}
  \centerline{\includegraphics[width=7cm, height=5cm]{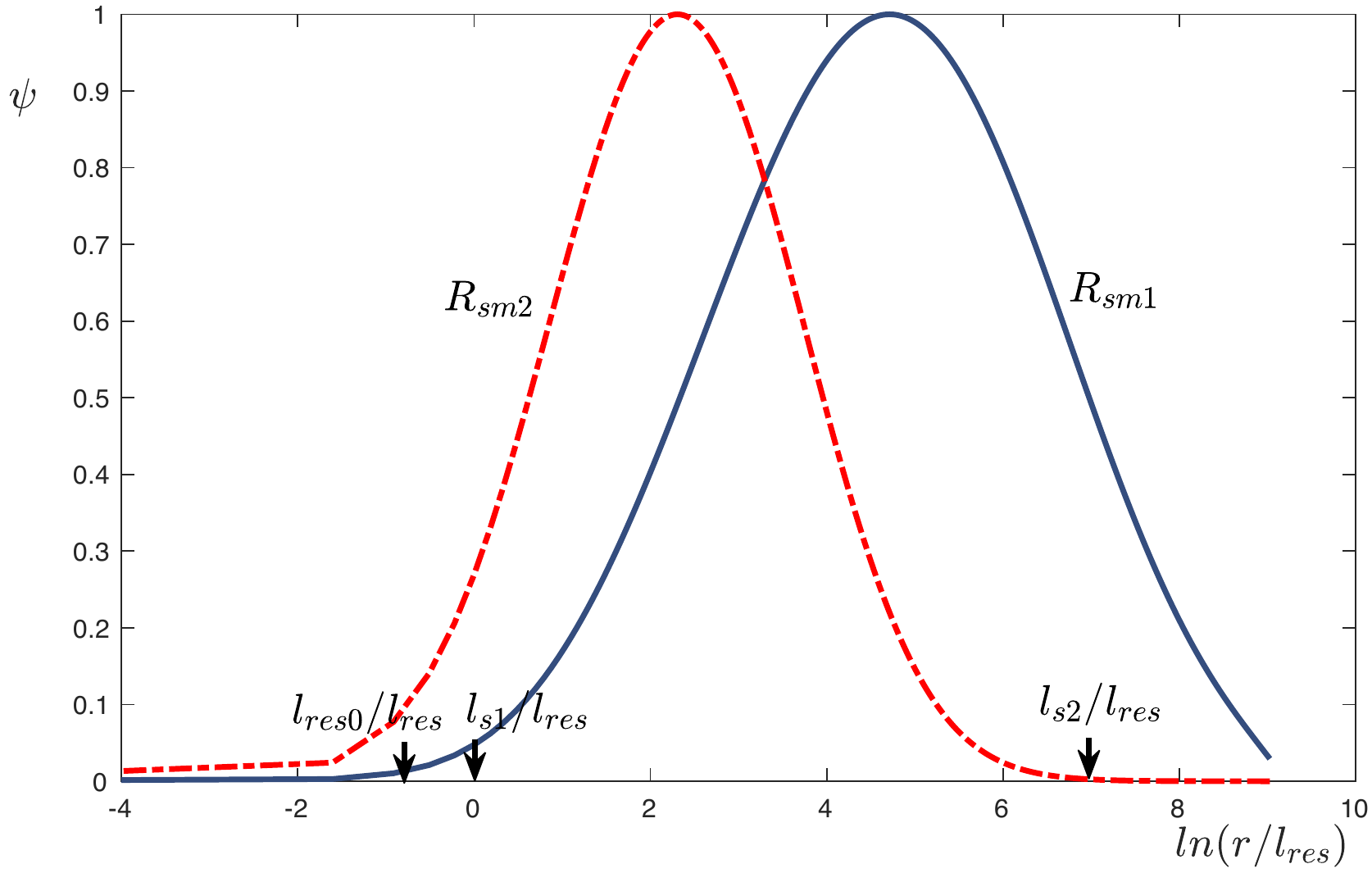}}% Images in 100% size
  \caption{The spatial structure of the fastest growing mode for two values of $R_{sm}$ for ${\cal C}=0.8$. $R_{sm1}=1$, $R_{sm2}=3.61\times 10^4$, $l_{s1}/l_{res}=1$, $l_{s2}/l_{res}=10^3$}
\label{fig:spatial}

\end{figure}

{\it Dimensionless form}. The characteristic eddy turnover time $\tau_r$ at scale $r$ is $D^{-1}r^{2-\xi}$. The requirement that that time decreases monotonously with $r$ yields the constraint $\xi<2$ and $\xi_0<2$ (the case of $\xi=2$ that characterises smooth flow is not considered here). The scale $l_s$ is fixed by requiring that the characteristic velocity $r/\tau_r$ at $r=l_s$ equals the speed of sound.
We use $l_{s}$ as the unit of length and the eddy turnover time at the sonic scale $\tau_s\equiv D^{-1}l_s^{2-\xi}=D_0^{-1}l_s^{2-\xi_0}$ as the unit of time. Then Eq.~(\ref{schrodinger}) can be written as,
\begin{equation}
\partial _t{\hat \psi}=2\bigl [R_{sm}^{-1}+{\cal D}\bigl ( 1+{\cal P}{\cal X}\bigr)\bigr]\partial^2_r {\hat \psi}-\tau_{s} U(rl_s){\hat \psi},
\label{nondimensionaleq}
\end{equation}
where ${\cal D}=r^{\xi_0}\!+\!(r^{\xi}\!-\!r^{\xi _0})\tanh r$, ${\cal P}={\cal C}\tanh r$, ${\cal X}=\xi_0\!+\!(\xi\!-\!\xi_0)\tanh r$, $R_{sm}\equiv l_s^2/\eta \tau _s$ is the sonic magnetic Reynolds number, and for simplicity the notations of the dimensionless independent variables have been left unchanged.
%introduce the magnetic resistivity scale $l_{res}=(\eta/D_0)^{1/\xi_0}$ defined with parameters of the flow below $l_s$. We use $l_{res}$ as the unit of length and the resistive time $\tau_{res}=l^2_{res}/\eta$ as the unit of time. Then Eq.~(\ref{schrodinger}) becomes $\partial _t{\hat \psi}\!=\!\left(2\!+\!S_L(r l_{res})/\eta\right)\partial^2_r {\hat \psi}\!-\!\tau_{res} U(r l_{res}){\hat \psi}$ that can be written as,
%\begin{eqnarray}&&\!\!\!\!\!\!\!\!\!\!\!\!\!
%\partial _t{\hat \psi}\!=\!%\left(2\!+\!\frac{S_L(r l_{res})}{\eta}\right)\partial^2_r {\hat \psi}\!-\!\tau_{res} U(r l_{res}){\hat \psi}=
%2\partial^2_r {\hat \psi}
%%\\&&\!\!\!\!\!\!\!\!\!\!\!\!\!
%\left(1 \!+\!\left(r^{\xi _0}\!+\!\left(S^{\xi_0-\xi}r^{\xi}\!-\!r^{\xi _0}\right)\tanh \left(\frac{r}{S}\right)\right)
%\right.\nonumber\\&&\!\!\!\!\!\!\!\!\!\!\!\!\!\left.
%(1\!+\!{\cal C}\tanh \left(\frac{r}{S}\right)\left(\xi_0\!+\!(\xi\!-\!\xi_0)\tanh \left(\frac{r}{S}\right)\right)\right)
%\nonumber\\&&\!\!\!\!\!\!\!\!\!\!\!\!\!
%-\tau_{res} U(r l_{res}){\hat \psi}
%,\label{dmless}
%\end{eqnarray}
%where we used the same variables for dimensionless coordinate and time and introduced the dimensionless sonic resistivity $S\equiv l_s/l_{res}$.
%We have,
%\begin{eqnarray}&&
%\tau_{res} U(r l_{res}){\hat \psi}=
%\end{eqnarray}
The faster decay of the spectrum in the supersonic inertial range than in the inertial range of incompressible turbulence implies that $\xi\geq \xi_0$.

{\it Results.} Inserting ${\hat \psi (r,t)}=\psi (r)e^{-\gamma t}$ into Eq. (\ref{nondimensionaleq}), we solve the resulting eigenvalue problem numerically. We define the resistive scale $l_{res}=(\eta/D)^{1/\xi}$
and resistive time $\tau_{res}=1/\gamma _{res}=l^2_{res}/\eta$ by the flow properties above $l_s$ assuming that $l_{res}\geq l_s$.
Defining further  $S \equiv l_s/l_{res}$ the sonic magnetic Reynolds number is given by $R_{sm}=S^{\xi}$. Therefore, if $R_{sm} \ll 1$ then the resistive scale is deep inside the supersonic inertial range. Scales below $l_s$ are then irrelevant and the framework of \cite{fos} applies and the characteristic growth time is $\tau _{res}$.  In contrast, for $R_{sm}\gg 1$ the growth rate scales with $\tau _{res0}^{-1}=l_{res0}^2/\eta$ where $l_{res0}=(\eta/D)^{1/\xi_0}= S^{-\xi/\xi_0} l_s$ is the resistive scale determined by the subsonic inertial range.

The dependence of the growth rate $\gamma$ of the fastest growing mode on the sonic magnetic Reynolds number $R_{sm}$ is depicted by the blue (full) curve in FIG. \ref{fig:gamma}.  The red (dashed) line represents the solution of eq. \ref{nondimensionaleq} with ${\cal D}=r^{\xi}$, ${\cal P}={\cal C}$, and ${\cal X}=\xi$, namely, not taking into account the presence of the sonic scale. The latter, denoted as the compressible limit line,  is a straight line with slope $1/3$ (namely $|\gamma |\tau _s =R_{sm}^{1/3}$) which signifies the fact that in that case the growth rate scales with $\tau _{res}^{-1}$.  For small values of $R_{sm}$ the curve coincides with the compressible limit line.  Indeed, as $R_{sm}$ decreases, larger portions of the turbulent domain reside in the compressible regime and the growth rate tends to the limit that is characterised by the compressible resistive time scale $\tau _{res}$, as in \cite{fos}. In the opposite limit, as $R_{sm}$ is increased, higher scales of the turbulence cross over into the incompressible range. The curve then departs from the compressible limit line and tends asymptotically to a straight line with slope $1/2$. This reflects the fact that in that regime the growth rate is characterised by the incompressible resistive time $\tau_ {res0}$.

The spatial structure of the fastest growing mode is depicted in FIG. \ref{fig:spatial} for $R_{sm}=1$ (full curve) as well as for $R_{sm}=3.61\times 10^3$ (dot-dash curve line). As may be seen, increasing  $R_{sm}$ results in shifting the amplified magnetic fields to lower scales as well as to more highly peaked spatial scale distributions.  The width of the eigenfunctions is quite large: $\sim250l_{res}$ for $R_{sm1}$ and $\sim50l_{res}$ for $R_{sm2}$, while the location of the maximum is about order of magnitude larger than $l_{res}$ for $R_{sm2}$ and two orders of magnitude larger $l_{res}$ for $R_{sm1}$.  As may be further noticed from the location of $l_s/l_{res}$ in both cases, the magnetic field amplification for low values of $R_{sm}$ occur at the supersonic region of the inertial range while for high values of the sonic magnetic Reynolds number the amplified magnetic field is concentrated in the smaller subsonic scales.

%%%%%%%%%%%%%%%%%%%%%%%%%%%%%%%%%%%%%%%%%%%%%

%%%%%%%%%%%%%%%%%%%%%%%%%%%%%%%%%%%%%%%%%%%%%

\end{document}